\documentclass[12pt,preprint]{aastex}
\begin{document}

\title{Discovery of a nearby twin of SN1987A's nebula around the
luminous blue variable HD168625: Was Sk--69$\arcdeg$202 an
LBV?\altaffilmark{1}}

\author{Nathan Smith}
\affil{Astronomy Department, University of California, 601 Campbell
Hall, Berkeley, CA 94720; nathans@astro.berkeley.edu}

\altaffiltext{1}{Based on observations made with the Spitzer Space
Telescope, which is operated by the Jet Propulsion Laboratory,
California Institute of Technology, under NASA contract 1407.}

\begin{abstract}

{\it Spitzer} images of the luminous blue variable (LBV) candidate
HD168625 reveal the existence of a bipolar nebula several times larger
than its previously-known equatorial dust torus.  The outer nebula of
HD168625 has a full extent of $\sim$80\arcsec\ or 0.85 pc, and one of
the lobes has a well-defined polar ring.  The nebula is a near twin of
the triple-ring system around SN1987A.  Because of these polar rings,
and accounting for stellar/progenitor luminosity, HD168625 is an even
closer twin of SN1987A than the B supergiant Sher~25 in NGC3603.
HD168625's nebula was probably ejected during a giant LBV eruption and
not during a red supergiant phase, so its similarity to the nebula
around SN1987A may open new possibilities for the creation of
SN1987A's rings.  Namely, the hypothesis that Sk--69$\arcdeg$202
suffered an LBV-like eruption would avert the complete surrender of
single star models for its bipolar nebula by offering an alternative
to an unlikely binary merger scenario.  It also hints that LBVs are
the likely progenitors of some type II supernovae, and that HD168625's
nebula is a good example of a pre-explosion environment.


\end{abstract}

\keywords{circumstellar matter --- stars: evolution --- stars:
  individual (HD168625) --- stars: mass loss --- stars: winds,
  outflows --- supernovae: individual (SN1987A)}

\section{INTRODUCTION}

Massive stars expel most of their initial mass on their journey from
the beginning of the main sequence to the Wolf-Rayet phase.  While
steady line-driven winds may contribute some of the envelope shedding,
much of the mass loss probably occurs near the end of core-H burning
in a series of continuum-driven eruptions or explosions (Smith \&
Owocki 2006).  These rarely-observed events (see Humphreys et al.\
1999) occur during a brief hot supergiant phase when the star is seen
as a luminous blue variable (LBV; Humphreys \& Davidson 1994).
Observations of expanding circumstellar nebulae around LBVs indicate
that masses of a few tenths to a few tens of solar masses can be
ejected in a single giant eruption that lasts only a few years (Smith
\& Owocki 2006; Clark et al.\ 2005; Smith et al.\ 2003).  The trigger
and energy supply for these blasts still evades understanding, but it
is expected that they may recur on timescales of several hundred
years.

The potential recurring nature of these outbursts provides a fairly
obvious observational prediction: some LBVs should be surrounded by
multiple nested ejecta shells from a sequence of previous outbursts.
LBVs are extremely rare and the detection of faint filamentary shells
is hampered by the bright central stars.  However, there are a few
well-documented cases where ancient filamentary shells with ages of a
few thousand years have been detected outside of the younger LBV
nebulae, such as $\eta$ Carinae (Walborn 1976; Smith et al.\ 2005;
Bohigas et al.\ 2000), P Cygni (Meaburn 2001; Meaburn et al.\ 1996,
1999, 2004), and HR Carinae (Nota et al.\ 1997; Weis et al.\ 1997).
This paper reports the new discovery of another example in an outer
shell around HD168625.

HD168625 (Hen 3-1681) is an 8th magnitude mid-B supergiant found at
the outskirts of M17, only $\sim$1\arcmin\ away from the LBV star
HD168607 (Chentsov \& Luud 1989).  It is probably located at a
distance of $\sim$2.2 kpc (van Gendren et al.\ 1992), although
distances of 1.2--2.8 kpc have been proposed (Robberto \& Herbst 1998;
Pasquali et al.\ 2002).  Chentsov \& Gorda (2004) have argued
convincingly that HD168625 and HD168607 are close to one another in
space, and that both are part of Sgr OB1 (Humphreys 1978) at 2.2
kpc.\footnote{If true, the pair HD168625 and HD168607 is just one
example of many pairs of eerily similar massive stars that are closely
spaced on the sky (e.g., Walborn \& Fitzpatrick 2000).}  Its bright,
dusty shell nebula was discovered by Hutsem\'{e}kers et al.\ (1994),
and has since been studied further by Nota et al.\ (1996), Robberto \&
Herbst (1998), Pasquali et al.\ (2002), and O'Hara et al.\ (2003).
The 12\arcsec$\times$16\arcsec\ ring nebula is expanding at about 20
km s$^{-1}$, has a dynamical age of a few thousand years, is N
enriched, and has an estimated total mass of 0.2--2 M$_{\odot}$ (the
lower estimates come from dust mass $\times$ 100, while the higher
estimates come from direct estimates of the gas mass).  The atmosphere
of the central star also has nitrogen-enhanced composition
(Garcia-Lario et al.\ 2001). Like other LBVs, HD168625's nebula has
bright infrared [Fe~{\sc ii}] emission, indicating a high-density,
low-ionization shell (Smith 2002).  The newly discovered nebula
reported here is far outside this shell.

Strictly speaking, the blue supergiant HD168625 is still a
``candidate'' LBV because it has not yet been observed to exhibit the
classical LBV or S Doradus-type photometric variability (Sterken et
al.\ 1999; van Genderen 2001; Humphreys \& Davidson 1994).  However,
it is widely suspected that the presence of a dense shell nebula like
that around HD168625 points toward a previous giant LBV eruption; van
Genderen (2001) classifies HD168625 as an ex/dormant LBV for this
reason.

The discovery of the ancient outer nebula around HD168625 reported
here is more significant than just one more example added to the three
nested LBV nebulae mentioned above, due to the relatively low
luminosity of this LBV.  On the HR diagram, it sits at the bottom of
the range of luminosities for currently-known LBVs (see Smith, Vink,
\& de Koter 2004).  This low luminosity means that it may have already
passed through a red supergiant (RSG) phase, and makes it of great
interest for examining the full parameter space for mass loss through
violent LBV outbursts.  As discussed in more detail below, its similar
morphology to the nebula around SN1987A makes it particularly
interesting.

\section{OBSERVATIONS}

{\it Spitzer} images of HD168625 were obtained with the Infrared Array
Camera (IRAC; Fazio et al.\ 2004) as part of the GLIMPSE survey
(Benjamin et al.\ 2003) of the inner Milky Way.  The reader is
referred to the GLIMPSE team website for details about the data
acquisition and processing.\footnote{see {\tt
http://www.astro.wisc.edu/sirtf/}.}  IRAC data for the region around
the target were retrieved from the archive through the GLIMPSE team
website, where the full image set was released on 2006 September 1.
The resulting {\it Spitzer} images in the four IRAC bands at 3.6, 4.5,
5.8, and 8.0~$\micron$ are displayed in Figure 1.  The insets in
Figures 1$a$, $b$, and $c$ show a different intensity scale in the
central region including the circumstellar dust ring that has already
been studied at mid-IR wavelengths using ground-based data with higher
spatial resolution (Robberto \& Herbst 1998; Meixner et al.\ 1999;
O'Hara et al.\ 2003).  The central star saturated the IRAC detector in
the 3.6 and 4.5~$\micron$ images, while warm circumstellar dust within
$\sim$10\arcsec\ of the star badly saturated the detector at
8.0~$\micron$.  The nearly vertical streaks in Figure 1 at
P.A.=--5\arcdeg\ from both HD168625 and 168607, as well as what appear
to be nearby companion stars along this same P.A.\ are both detector
artifacts caused by the bright central stars.

The previously-known 12\arcsec$\times$16\arcsec\ ring nebula around
HD168625 is clearly detected in all four IRAC bands in Figure 1; in
fact, the whole ring is saturated in the 8.0~$\micron$ image.  The
8.0~$\micron$ image in Figure 1$d$ also shows more extended nebulosity
out to $\sim$40\arcsec\ from the star.  The nebula is apparently
bipolar, with a thin loop or ring to the northeast (NE) and some
emission with similar extent to the SW.  The presence of this extended
emission is dubious at 3.6 and 5.8 $\micron$, but is clearly seen at
4.5 $\micron$.  This extended structure is real, since IRAC images
reveal no comparable extended features around the neighboring star
HD168607.  This is the case at visual wavelengths as well
(Hutsem\'{e}kers et al.\ 1994).  In hindsight, some H$\alpha$ emission
can be seen coming from outer regions of the nebula in Figure 1 of
Hutsem\'{e}kers et al.\ (1994).  However, there is no indication that
this diffuse H$\alpha$ emission is related to HD168625, so
Hutsem\'{e}kers et al.\ attributed it to the background H~{\sc ii}
region M17.

Figure 2 shows the sky-subtracted spectral energy distribution (SED)
of the surface brightness at a position in the NE polar ring.  A 370~K
Planck function with $\lambda^{-1}$ emissivity is shown for comparison
with the data.  However, this is of limited use for inferring physical
properties of the nebulosity, since scattered light or emission
features such as Br$\alpha$, polycyclic aromatic hydrocarbons, or
silicate emission may contribute to the emission seen in the IRAC
filters.  Although dust at $\sim$370~K seems to fit the data fairly
well, it would be surprising to find dust at that high a temperature
so far from the central star.  At the likely separation of the NE ring
from the central star ($\sim$40\arcsec\ or 88,000 AU), dust heating by
the central engine is negligible, and the dust should be at a similar
temperature to cool grains in the surrounding interstellar medium.
Thus, it is likely that the IRAC SED is dominated by a combination of
scattered light, thermal dust continuum, and emission features at
these wavelengths, so deriving a reliable dust mass is not possible
here.  IR spectroscopy of the nebula would be useful, but it was not
observed with the SWS or LWS spectrographs on {\it ISO}, and HD168625
has not yet been observed with the IR spectrograph (IRS) on {\it
Spitzer}.  There is an interesting discrepancy in mid-to-far IR
photometry, though, in that fluxes from the {\it Infrared Astronomical
Satellite} ({\it IRAS}) are all systematically larger than fluxes
obtained with ISOPHOT, the {\it Midcourse Space Experiment} (MSX), or
ground-based mid-IR imaging, all of which have smaller beamsizes than
IRAS.  This hints that cool dust in the more extended outer nebula
discovered here may have substantial flux at longer wavelengths, so
submm data would be beneficial to measure or place limits on its mass.
Unfortunately, the 100 $\micron$ IRAS flux density of $\sim$580 Jy
listed in the IRAS point-source catalog is flagged as unreliable.

\section{RESULTS: MORPHOLOGY AND GEOMETRY}

\subsection{The Outer Rings and Bipolar Nebula}

{\it Spitzer}/IRAC images reveal a previously-unknown large bipolar
nebula around HD168625, extended 40\arcsec\ toward the NE and toward
the SW.  The most interesting feature is a well-defined ellipse or
ring in the NE polar lobe.  The discovery of such a ring around an
evolved hot supergiant star is of great interest with regard to its
implications for the nebula around SN1987A.  This is discussed further
in \S 4.3 and \S 4.4.

Figure 3 shows that the entire visible part of the NE ring matches a
smooth ellipse, corresponding to a circle inclined 61\arcdeg\ from the
plane of the sky, and rotated with its semi-minor axis (the polar
axis) at a position angle of 23\arcdeg.  The parameters of this
ellipse in Figure 3 are accurate to within $\la$5\%.  The ring's FWHM
thickness is about 2 pixels (2$\farcs$4), which matches the
diffraction limit of the {\it Spitzer} telescope at 8.0~$\micron$, so
the ring's thickness is unresolved in IRAC images.

The SE polar lobe does not have such a clearly-defined ring in IRAC
images.  The dashed ellipse drawn in Figure 3 is meant to illustrate
that a ring identical to that in the NE polar lobe can match the
extent of the diffuse 8.0~$\micron$ emission in the SW polar lobe.
Thus, it is plausible that such a ring defines the outer extremity of
a conical lobe.  A line connecting the centers of the NE and SW rings
is drawn in Figure 3 (representing the polar axis), and it passes
within 1\arcsec\ of the central star.  Figure 3 also shows a smaller
ellipse drawn in black over the inner equatorial ring.  This ellipse
has the same inclination, position angle, and polar axis, but the
radius is one-quarter of the NE polar ring.

Thus, the bipolar lobes appear to be axisymmetric in basic shape and
extent, but asymmetric in their detailed morphology.  Potential
sources of this asymmetry are 1) stronger ionization of the northern
lobe by UV radiation from stars in the nearby H~{\sc ii} region M17,
2) influence of the LBV wind from the nearby star HD168607 that may
have disturbed a hypothetical SW polar ring, or 3) intrinsic asymmetry
in the star's mass ejection toward each of its poles.  Whatever the
cause, it is interesting that only the northern half of the inner
ellipsoidal shell is seen in H$\alpha$ images (Hutsem\'{e}kers et al.\
1994).

The first option of preferential external UV illumination of
HD168625's nebula by M17 is quantitatively plausible if they are
located at the same distance.  HD168625 is projected on the sky about
11\arcmin\ from the ionizing star cluster at the center of M17,
indicating a probable separation of $\sim$15 pc (including a factor
$\sqrt{2}$).  If they really are at the same distance, HD168625 would
seem to have a suspiciously clear shot to see these O stars, as the
large-scale horseshoe or ``V'' shape of M17 opens directly toward
HD168625.  Hanson, Howarth, \& Conti (1997) estimated a total Lyman
continuum luminosity of 1.2$\times$10$^{50}$ s$^{-1}$ for the O stars
that power M17, indicating a Lyman continuum flux of
$\Phi_H\simeq$5$\times$10$^9$ s$^{-1}$ cm$^{-2}$ in the vicinity of
HD168625 for a separation of 15~pc.  With ionization balance given
roughly by $\Phi_H \simeq \alpha_{\rm B} l n_e^2$, this external
ionizing flux could support electron densities of 500--1000 cm$^{-3}$
if the typical path length $l$ is $\sim$1\arcsec\ or
3.3$\times$10$^{16}$ cm ($\alpha_{\rm B}$=2.3$\times$10$^{-13}$ cm$^3$
s$^{-1}$ is the case B recombination coefficient).  This is comparable
to typical densities estimated in similar LBV nebulae, and is
comparable to the expected ionizing flux from a mid-B supergiant at
the radius of the ring.  The second option of external influence from
the stellar wind ram pressure of the neighboring LBV HD168607 is
plausible simply because the two stars have similar mass-loss rates,
and the projected separations between the gas in the SW polar lobe and
each of the stars is comparable. In either case where external effects
cause the asymmetry, special geometric orientations are necessary so
that only one lobe feels the external influence.  The third
alternative, where the star's mass-loss geometry is intrinsically
asymmetric toward its two polar hemipsheres, is not easily
accomplished either and would be rare among known bipolar nebulae.

\subsection{The Inner Equatorial Ring}

Previous optical/IR imaging of the inner dust ring found it to be a
flattened toroid or ring, with a radius of $\sim$8\arcsec\ (0.085 pc)
and an inclination angle\footnote{Here inclination is defined as the
tilt of the equatorial plane from the plane of the sky, as in binary
systems.} of $i \simeq$60\arcdeg$\pm$15 (O'Hara et al.\ 2003; Pasquali
et al.\ 2002, Robberto \& Herbst 1998; Nota et al.\ 1996;
Hutsem\'{e}kers et al.\ 1994).  Observed kinematics showed that it had
a radial expansion speed of 19--20 km s$^{-1}$ (Hutsem\'{e}kers et
al.\ 1994; Pasquali et al.\ 2002), and consequently, a dynamical age
of a few thousand years.  This inner ring is taken to define the
equatorial plane.  Imaging also revealed a fainter ellipsoidal shell
that met the equtorial ring, but was more elongated along the polar
axis (e.g., Hutsem\'{e}kers et al.\ 1994; O'Hara et al.\ 2003; Nota et
al.\ 1996).  Emission from this ellipsoidal shell, which is brightest
in the NE polar lobe at radii up to about 12\arcsec\ from the star,
can be seen in the insets of Figure 3.

The structure of the inner ring in IRAC images seems to suggest a
slightly lower inclination (white ellipse in the inset of Figure 3)
than 61\arcdeg\ (black ellipse).  However, detailed study of IR and
optical images with higher spatial resolution found a best fit
inclination of 60\arcdeg$\pm$15\arcdeg, as noted above.  This is
within 1\arcdeg\ of the inclination derived here for the outer NE
polar ring.  Therefore, the outer polar rings and the inner equatorial
ring are plane-parallel.  There is a slight tilt in the nebula, in the
sense that the line connecting the centers of the two polar rings is
not perpendicular to the major axes of the rings.

\subsection{Geometric Interpretation}

Figure 4 shows a sketch of the proposed 3-D geometry of the ring
system and bipolar nebula around HD168625 as viewed from the western
side, near the equatorial plane (an Earth-based observer is to the
left).  The NE ring is a well-defined ellipse, providing an accurate
measure of several geometric parameters in the system under the
assumption that the object is a circular ring inclined to our line of
sight.  The parameters in Figure 4 and Table 1 are based on the
observed dimensions of the NE ring, and all assume a distance of 2.2
kpc.

The NE ring is at a latitude of 41\arcdeg\ from the equator, so half
the opening angle of the conical bipolar lobes would be 49\arcdeg.
The polar ring resides at a height above the equatorial plane of 0.3
pc, and a radial distance from the star of 0.46 pc.  The kinematics of
the outer polar rings have not yet been measured, but this is an
important observational goal. If they have the same $\sim$4,000 yr
dynamical age as the inner ring, they would be expanding at roughly
110 km s$^{-1}$.  This is still slower than the central star's
observed terminal wind velocity of $v_{\infty}$=183 km s$^{-1}$ (Nota
et al.\ 1996), so it is at least plausible that the inner and outer
rings are coeval.

The polar axis is tilted out of the plane of the sky by
90\arcdeg-$i$=29\arcdeg, but it is not certain from images alone which
way the polar axis is pointing.  Figure 4 is drawn with the NE polar
axis tilted toward us and the SW polar axis tilted away, such that the
expansion center of the NE ring would be blueshifted.  This is based
primarily on the observed morphology and kinematics of the
well-studied inner ring; the inner equatorial ring is brighter on its
southern rim in scattered light (Pasquali et al.\ 2002); predominantly
forward scattering of dust grains at visual wavelengths would imply
that this is the near side.  If the south/SW part of the equatorial
ring is closer to us, then the NE polar axis would need to be tilted
toward us.  This orientation is also favored by the observed expansion
velocities in the inner ring (Hutsem\'{e}kers et al.\ 1994; Pasquali
et al.\ 2002).  Further study with deep high-resolution spectra of the
outer nebula could provide a more certain answer.

{\it Spitzer} images also reveal extended diffuse emission just
outside the inner ring that was not revealed in previous IR imaging.
This may represent warm dust in the side walls of the biconical
surface of the polar lobes, which may connect the inner ring to the
outer rings, as depicted by the shaded regions in Figure 4.  These
side walls of the biconical lobes are drawn as slightly curved
surfaces in Figure 4, but this detail of the geometry remains to be
determined spectroscopically.  These thin walls have counterparts in
the ``sheets'' of material that seem to connect the inner and outer
rings of SN1987A (Burrows et al.\ 1995).

\section{DISCUSSION}

\subsection{Swept-Up RSG Wind or LBV Eruption?}

Although it may be plausible that HD168625 has already passed through
a RSG evolutionary phase before becoming an LBV, it is highly unlikely
that most of the mass in the outer nebula reported here was ejected in
that previous RSG phase and shaped afterward.  Doppler shifts or
proper motions of the outer nebula have not yet been measured, so its
dynamical age is unknown.  However, there are essentially two
possibilities worth considering: the outer bipolar nebula could have
been ejected at the same time as the inner ring, with a dynamical age
of $\sim$4,000 yr (Table 1), or it could have been ejected much
earlier.  (A scenario where the outer bipolar lobes are significantly
younger than the inner ring can be rejected because their required
expansion speed would be faster than the stellar wind's terminal
velocity.)  Both these two possibilities are inconsistent with a
nebular origin during the RSG phase, for the following reasons:

\begin{itemize}

\item If the inner and outer rings are coeval, then the required
expansion speed of the outer rings is V$_{\rm exp}\simeq$110 km s$^{-1}$
(Table 1).  This high speed rules out a RSG origin for the nebula.  On
the other hand, such speeds are typical in giant LBV eruptions.

\item If the inner and outer rings are {\it not} coeval, and the outer
rings were ejected much earlier, then a RSG origin is probably ruled
out as well.  This is because the creation of {\it multiple} shells in
an interacting RSG/LBV winds scenario would require the star to
transit from the RSG to the LBV phase twice in a few thousand years.
Two blue loops in the HR diagram are not predicted by current
evolutionary models (e.g., Arnett 1991; Maeder \& Meynet 2000;
Fitzpatrick \& Garmany 1990).  While the nebula's dynamical timescale
is in marginal agreement with a single transition episode from RSG to
blue supergiant, it is an order of magnitude shorter than the time
that would be needed for two full blue loops, set by the thermal
timescale of the He core.

\end{itemize}

Of course, the expectation that the nebula was most likely ejected in
one or more LBV eruptions does not preclude the possibility that
HD168625 has already passed through a RSG phase.  In fact, low
luminosity LBVs probably {\it are} post-RSGs, in order to raise their
L/M ratio enough that they are subject to the LBV instability (see
Smith et al.\ 2004; Humphreys \& Davidson 1994).  In any case, it
would be desirable to obtain high-resolution spectra of the outer
nebula around HD168625, in order to measure its apparent Doppler
velocities and chemical abundances.  Observed line-of-sight motion can
be deprojected with the geometry in Figure 4 to independently derive
the dynamical age of the outer nebulosity.

What causes the bipolar shape of the outer nebula?  No evidence for a
close companion star has been reported, even though this star has been
studied spectroscopically and photometrically in the interest of
testing its LBV nature (e.g., Chentsov \& Luud 1989; Sterken et al.\
1999; Nota et al.\ 1996; Pasquali et al.\ 2002).  At the high
inclination of the ring system ($i\simeq$61\arcdeg), we might expect
orbital reflex motion or even periodic eclipses to be detectable if
the central star is a close binary, so renewed monitoring of this star
may be interesting. Evidence of orbital reflex motion would go a long
way toward demonstating that tidal spin-up by a close companion may be
related to bipolar geometry.  Whether spun-up by a companion or not,
though, a star in near-critical rotation is probably the culprit (see
\S 4.2).  LBV eruptions from rotating stars should produce bipolar
mass loss (e.g., Owocki 2003), as demonstrated in the case of $\eta$
Carinae (Smith 2006).

\subsection{Implications for Mass Loss Through LBV Eruptions}

The dynamical age for the new outer bipolar nebula around HD168625 has
not yet been measured, but it is at least $\sim$4000 yr.  It is the
fourth Galactic LBV/LBVc to exhibit such a large outer filamentary
shell, following $\eta$~Carinae (Walborn 1976; Smith et al.\ 2005;
Bohigas et al.\ 2000), P~Cygni (Meaburn 2001; Meaburn et al.\ 1996,
1999, 2004), and HR~Carinae (Nota et al.\ 1997; Weis et al.\ 1997).

The chief reason that the new discovery of this ancient nebula around
HD168625 is interesting for research on LBVs is because of the star's
relatively low luminosity compared to other LBVs.  HD168625 is at the
bottom of the range of luminosities for LBVs and LBV candidates on the
HR diagram (see Smith, Vink, \& de Koter 2004).  It has not been
observed to suffer the same photometric variability as more luminous
bona fide LBVs.  Its nebula, however, indicates that violent mass loss
through repeated giant LBV eruptions may be important across the full
range of LBV luminosity.  Furthermore, its bipolar morphology with a
tightly-pinched waist and no evidence for binarity in the central star
is consistent with the idea that rotation can drive bipolar mass-loss
(Owocki 2003; Dwarkadas \& Owocki 2002; Owocki et al.\ 1996; Cranmer
\& Owocki 1995), even at relatively low LBV luminosities.  Presumably
this can occur because the stars approach or violate the classical
Eddington limit during their giant eruptions when the mass is ejected,
allowing the star's rotation to be more influential at the resulting
lower effective gravity (e.g., Langer 1998; Glatzel 1998; Maeder \&
Meynet 2000).

\subsection{A Galactic Analog of the Nebula Around SN1987A}

The multi-ringed geometry of HD168625's nebula depicted in Figure 4
makes it the Milky Way's closest analog of the famous triple-ring
nebula around SN1987A.  The two nebulae are compared in Table 1, where
we see that all of HD168625's parameters are very close to those of
SN1987A.  When a range of parameters is given, they often overlap.
Both have nitrogen-enriched nebulae, although the level of
N-enrichment in HD168625 is not as well constrained (see Table 1).
The only substantive physical difference between the two nebulae (more
than $\pm$30\%) is that HD168625's inner equatorial ring is smaller
and its dynamical age is less.  Given more time, the stellar wind of
HD168625 ($\sim$183 km s$^{-1}$) will continue to sweep up its
equatorial ring to have a larger radius more akin to that of SN1987A,
Rayleigh-Taylor instabilities will lead to protrusions like the
``fingers'' that cause the hot spots in SN1987A's ring (e.g., Michael
et al.\ 2000), and the ring's expansion speed may decrease as the wind
sweeps up more mass.  At present, the outer nebula around HD168625 is
very difficult to see at visual wavelengths because it is overwhelmed
by light from the bright central star, and the nebula may be mostly
neutral with only weak intrinsic H$\alpha$ emission.  However, it is
easy to imagine that in the not too distant future, an event may occur
that will flash-ionize HD168625's nebula and will remove the bright
central star, making the nebula appear identical to SN1987A's.

A comparison to SN1987A has also been made for the blue supergiant
Sher~25 in NGC3603 (Brandner et al.\ 1997a, 1997b), which harbors an
equatorial ring.  However, unlike HD168625, Sher~25 shows no sign of
polar rings, but instead shows irregular structure in the bipolar
lobes.  A few objects do show pairs of plane-parallel rings, like the
planetary nebula Abel 14 and the massive eclipsing binary RY Scuti
(Smith et al.\ 2002), but in these systems the ring pairs are at low
latitudes near the equator and they lack a third inner ring.  Many
planetary nebulae have an equatorial torus or ring at the waist of
their bipolar lobes (e.g., Balick \& Frank 2002), and with limb
brightening, some of these polar lobes may appear ring-like.  However,
HD168625 is the only one around a massive star, so among objects known
to date, its nebula may be the best analog of the pre-SN environment
around SN1987A.

Stellar properties further justify the close comparison between the
nebulae around HD168625 and SN1987A.  SN1987A's progenitor was
identified as the blue supergiant Sanduleak --69$\arcdeg$202 (Walborn
et al.\ 1987).  Its B3 Ia spectral type would correspond to T$_{\rm
eff}\simeq$16,000 K (Crowther et al.\ 2006), and its luminosity of
$\ga$10$^5$ L$_{\odot}$ would indicate an initial mass of roughly 20
M$_{\odot}$ (Arnett 1991; Woosley et al.\ 1987).  HD168625 is almost
identical by comparison --- it has T$_{\rm eff}\simeq$15,000 K and a
luminosity of 10$^5$ to 10$^{5.4}$ L$_{\odot}$ (Nota et al.\ 1996),
corresponding to a likely initial mass of 20--25 M$_{\odot}$.  Both
stars have probably passed through the RSG phase in the recent past
(see Smith et al.\ 2004).  On the other hand, Sher~25's parameters are
significantly different.  It's spectral type of B1.5 Ia indicates that
it is hotter than Sk --69$\arcdeg$202, and its luminosity of
10$^{5.9}$ L$_{\odot}$ (Smartt et al.\ 2002) is much higher.  In fact,
this luminosity is so high that it precludes Sher~25 from having been
a RSG (Humphreys \& Davidson 1994), so its evolution on the HR diagram
has been very different from Sk --69$\arcdeg$202.  Chemical abundances
reinforce this conjecture; Smartt et al.\ (2002) find that Sher~25's
N/O ratio is incompatible with a previous RSG phase.

\subsection{Potential Implications for the Progenitor of SN1987A}

The profound similarities discussed above beg the question: {\it Was
SN1987A's progenitor an unrecognized or quiescent LBV?}  If the nebula
had been discovered prior to 1987, then Sk --69$\arcdeg$202 would
likely have been classified as a low-luminosity LBV candidate by
today's criteria.  In a sense, an LBV hypothesis for the SN1987A
progenitor is a small modification from the established fact that it
was a blue supergiant, since the duration of the LBV phase one infers
from observed statistics is identical to the expected timescale for a
post-RSG blue loop at this stellar mass (e.g., Martin \& Arnett 1995).
The LBV hypothesis is also consistent with the nitrogen enrichment in
SN1987A's circumstellar ejecta (Fransson et al.\ 1989; Sonneborn et
al.\ 1997), since LBV nebulae are commonly N-rich (e.g., Smith \&
Morse 2004; Davidson et al.\ 1986; Lamers et al.\ 2001).  Furthermore,
there are growing hints that some small fraction of type II SN
progenitors are LBVs (e.g., Gal-Yam et al.\ 2006; Kotak \& Vink 2006;
Smith \& Owocki 2006; Smith 2007).

So then, {\it was the bipolar circumstellar nebula around SN1987A
ejected in an episodic eruption like those seen in LBVs, rather than
arising from a swept-up asymmetric RSG wind?}  There is evidence for
much larger-scale bipolar nebulosity well outside the triple rings
around SN1987A, seen in continuum emission from light echoes (e.g.,
Sugerman et al.\ 2005; Crotts et al.\ 1995; Wampler et al.\ 1990).
Formation of these outer bipolar lobes by interacting winds would
require more than one blue loop, but multiple blue loops are not
expected.  On the other hand, multiple nested nebulae with the same
geometry are expected from repeated LBV-like ejection events.  Here we
have seen that SN1987A's two closest Galactic analogs -- HD168625 and
Sher~25 -- both ejected their ringed bipolar nebulae as blue
supergiants, most likely in LBV outbursts (see \S 4.1).  Detailed
studies of $\eta$ Carinae (e.g., Smith 2006) have established a clear
precedent that near-critically rotating LBVs can eject bipolar nebulae
without resorting to a pre-existing equatorial density enhancement,
and that they can do so repeatedly with the same geometry.  In $\eta$
Car, it was even found that the mass concentration peaked at mid
latitudes of around 50--60\arcdeg\ (Smith 2006), similar to the
latitudes of the polar rings around HD168625 and SN1987A.  Thus, the
bipolar shape of LBV nebulae in general -- and the ringed structure of
HD168625's nebula in particular -- argue that a past LBV-like episode
may be an alternative mechanism to explain the formation of SN1987A's
rings.  The LBV hypothesis for SN1987A is appealing because it may
avert the complete surrender of rotating single star models as an
alternative to an improbable binary merger event.

Perhaps the strongest objection to this LBV progenitor scenario for
SN1987A would seem to be the low expansion speed of SN1987A's
equatorial ring, which is slow even compared to wind speeds of RSGs.
However, this is not much of a show stopper for the LBV hypothesis.
LBV nebulae exhibit a wide range of expansion speeds, sometimes in the
same object.  For example, $\eta$ Car's ejecta move at speeds from as
high as 600 km s$^{-1}$ or more down to speeds as low as 40 km
s$^{-1}$ (Smith 2006).  We might expect some very low speeds, since
the escape speed formally tends toward zero when a star violates the
classical Eddington limit during an LBV eruption (e.g., Zethson et
al.\ 1999).  Indeed, V$_{\rm exp}$ for HD168625's equatorial ring is
only 19 km s$^{-1}$, which is even slower than the 26 km s$^{-1}$
expansion speed of SN1987A's polar rings (recall that the fast polar
ring speed of 110 km s$^{-1}$ listed in Table 1 {\it assumes} that the
rings are all coeval).  Incidentally, HD168625's equatorial and
(assumed) polar ring expansion speeds are nearly identical to Sher
25's equatorial and polar speeds (Brandner et al.\ 1997a,
1997b). Another potential objection is the relatively low luminosity
of Sk--60$\arcdeg$202 compared to LBVs, but the low end of the
luminosity range for LBVs is not well characterized.  While it seems
unlikely the Sk--69$\arcdeg$202 could have been a bona-fide LBV
exhibiting classical S~Doradus variability at its relatively low
luminosity, it may nevertheless have suffered episodic mass loss
analogous to a giant LBV-type eruption.

The recognition that SN1987A's progenitor was a blue supergiant
(Walborn et al.\ 1987) instead of a RSG was surprising at the time,
and led to drastic revisions in our understanding of stellar evolution
for massive stars (e.g., Arnett 1991).  Furthermore, the high degree
of axial symmetry in the bipolar circumstellar nebula, thought to have
been ejected during the RSG phase, challenged our understanding of the
formation of bipolar nebulae (Blondin \& Lundqvist 1993; Martin \&
Arnett 1995).  It required an external source of angular momentum,
ultimately strengthening the case for a binary merger before the SN
event (Podsiadlowski 1992; Morris \& Podsiadlowski 2006; Collins et
al.\ 1999).  It is amusing to ponder how the debate about SN1987A's
pre-SN evolution may have unfolded somewhat differently, had it been
postulated at the time that the progenitor could have been a
low-luminosity quiescent LBV, or that the bipolar geometry originated
in an LBV-like eruption instead of in the RSG phase.

With the close comparison to SN1987A, an even more amusing notion is
the possibility that HD168625 will suffer a core collapse in the near
future.  Given its relatively small distance of 2.2 kpc, its low
interstellar extinction, and its declination of only --16\arcdeg, the
SN would be seen from the northern hemisphere and would likely be one
of the most spectacular celestial events in recorded history.

\acknowledgments  

This work was based on archival data obtained with the {\it Spitzer
Space Telescope}, which is operated by the Jet Propulsion Laboratory,
California Institute of Technology, under a contract from NASA. I used
images obtained as part of the GLIMPSE legacy science program, with
thanks to P.I.\ Ed Churchwell and members of the GLIMPSE team.  I also
thank Stephen Smartt, Dave Arnett, and an anonymous referee for
helpful discussions and comments on the manuscript.


\begin{deluxetable}{lccc}\tabletypesize{\scriptsize}
\tablecaption{Geometric parameters of the circumstellar rings}
\tablewidth{0pt}
\tablehead{
 \colhead{Parameter} &\colhead{units} &\colhead{HD168625} &\colhead{SN1987A} }
\startdata

inner ring inclination	&deg.		&61$\pm$2	&43$\pm$3	\\
outer ring inclination	&deg.		&60$\pm$15	&47--51		\\
outer ring thickness	&pc		&$<$0.026	&$<$0.08	\\
outer ring spherical radius 	&pc	&0.46		&0.5--0.64	\\
outer ring cylindrical radius R	&pc	&0.35		&0.43--0.45	\\
R(pole)/R(eq)		&\nodata	&4		&2.2		\\
ring height		&pc		&0.3		&0.3--0.43	\\
ring latitude		&deg.		&41$\pm$2	&32--45		\\
cone opening angle	&deg.		&98$\pm$2	&90--116	\\
P.A. of polar axis	&deg.		&23		&--9		\\

V$_{exp}$(pole)		&km s$^{-1}$	&(110)		&26		\\
V$_{exp}$(eq)		&km s$^{-1}$	&19		&8.3		\\
dynamical Age		&10$^4$ yr	&0.4		&2.2		\\

N/O			&(N/O)$_{\odot}$ &$>$3		&$\ga$10	\\

\enddata

\tablecomments{All values for HD168625 assume a distance of D=2.2
kpc.  V$_{exp}$(pole) is in parentheses because it has not been
measured, but was calculated assuming that it has the same age as the
inner ring.}

\tablerefs{Burrows et al.\ (1995); Plait et al.\ (1995); Crotts \&
Heathcote (1991); Meaburn et al.\ 1995; Fransson et al.\ (1989);
Sonneborn et al.\ (1997); Hutsem\'{e}kers et al.\ (1994).}

\end{deluxetable}

\begin{figure}
\epsscale{0.95}
\plotone{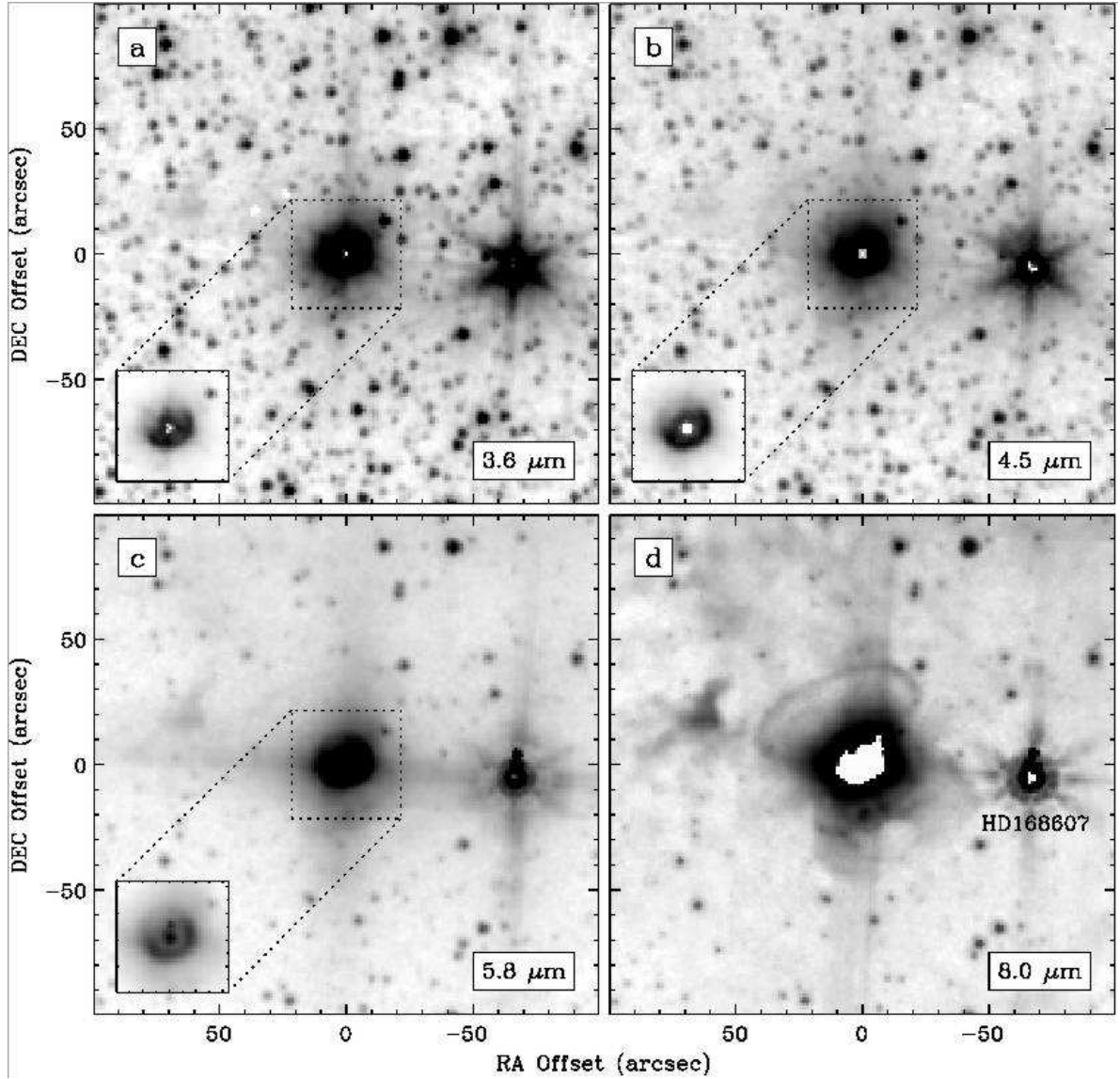}
\caption{{\it Spitzer} images of the nebula around HD168625 in the
four IRAC bands, taken as part of the GLIMPSE project.  The faint
outer bipolar nebula (especially the NE polar ring) is best seen at
8.0~$\micron$.  The blob located at offset +60\arcsec,+20\arcsec\ may
or may not be associated with the circumstellar ejecta.  HD168607 is
located $\sim$65\arcsec\ west of center.}
\end{figure}

\begin{figure}
\epsscale{0.45}
\plotone{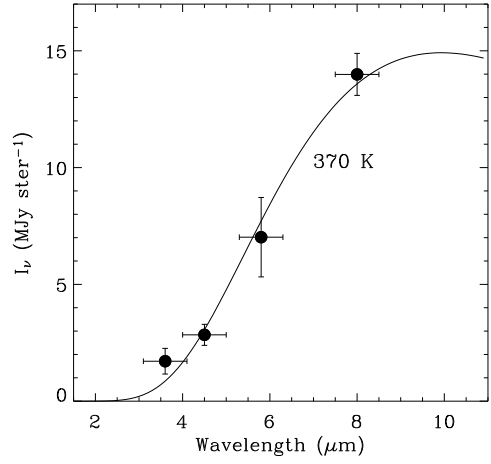}
\caption{The Spectral energy distribution of the surface brightness in
the NE polar ring.  The surface brightness was measured in a
2$\farcs$4 square aperture located 29\arcsec\ east and 28\arcsec\
north of HD168625, and the surface brightness of the nearby
background sky was subtracted.}
\end{figure}

\begin{figure}
\epsscale{0.45}
\plotone{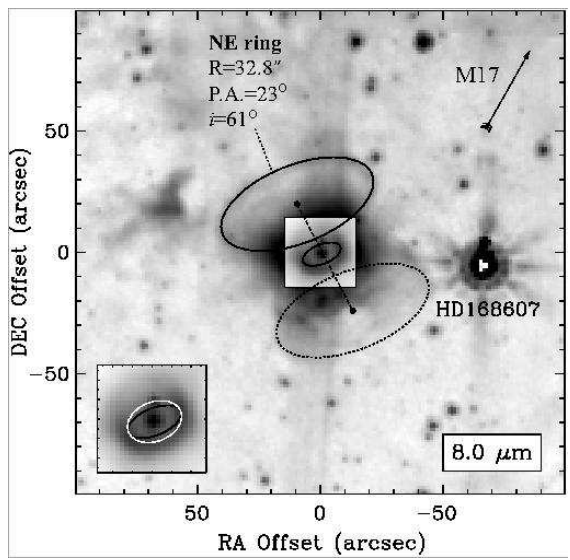}
\caption{The 8.0~$\micron$ image of HD168625 from Fig.\ 1$d$ with
representative rings superposed (for the saturated area around the
central star, the 5.8~$\micron$ image is substituted).  The NE ring
that is drawn here matches the size and shape of the observed emission
from the nebular ring.  The emission from the SE polar lobe is not as
well-defined as the NE ring, so drawn here is an ellipse of the same
size, aspect ratio, and position angle that is matched to the extent
of the polar lobe emission.  A dashed line connects the centers of
these two ellipses, representing a projection of the polar axis.  The
smaller ellipse in black superposed on the inner equatorial ring has
25\% of the size of the polar rings, with the same position angle and
inclination.  The inset in the lower left shows that an ellipse of the
same radius but with a slightly smaller inclination angle (white) fits
the inner ring better.  The arrow in the upper right indicates the
direction toward the center of the nearby H~{\sc ii} region M17.}
\end{figure}

\begin{figure}
\epsscale{0.45}
\plotone{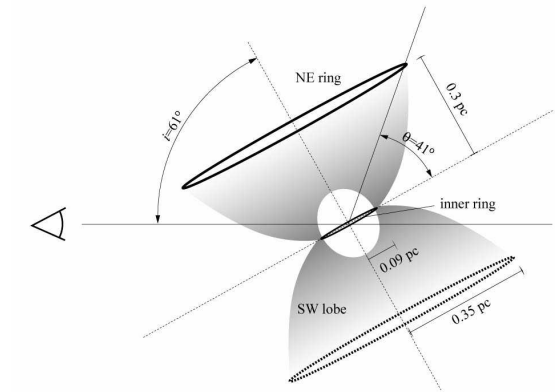}
\caption{A sketch of the proposed 3D geometry in the nebula around
HD168625, as viewed from the side, at a latitude near the equator.  An
Earth-based observer is at the left.  See Table 1 for details.}
\end{figure}

\end{document}